\documentclass[10pt,twocolumn,showpacs,preprintnumbers,amsmath,amssymb,aps,prb,longbibliography,superscriptaddress]{revtex4-2}
\usepackage{array,braket,mathtools,siunitx,bbm,enumitem}
\usepackage[unicode=true,
bookmarks=true,bookmarksnumbered=true,bookmarksopen=false,
breaklinks=true,pdfborder={0 0 0},backref=false,colorlinks=true]{hyperref}
\hypersetup{citecolor={blue},urlcolor={magenta}}
\usepackage{cleveref}
\crefname{appendix}{App.}{Apps.}
\crefname{equation}{Eq.}{Eqs.}
\crefname{figure}{Fig.}{Figs.}

\begin{document}
\title{Bootstrapping Quantum Hamiltonians with Symmetry}
\author{Michael G. Scheer}
\affiliation{Department of Physics, Princeton University, Princeton, New Jersey 08544, USA}
\affiliation{Department of Physics, Harvard University, Cambridge, Massachusetts 02138, USA}
\date{\today}

\begin{abstract}
We describe a semidefinite relaxation method which finds lower bounds to the ground state energy of a quantum Hamiltonian subject to Hermitian linear constraints along with approximations of ground state expectation values. We show that symmetry can be used to significantly reduce the computational requirements, and we include unitary, antiunitary, discrete, and continuous symmetries in our analysis. We demonstrate our method using the 1D Hubbard model and find quantitative agreement with both exact diagonalization and the Bethe ansatz.
\end{abstract}

\maketitle

\emph{Introduction}---Computing ground state properties of strongly interacting quantum many-body systems is a central problem in quantum chemistry, condensed matter theory, and high energy theory. Since the Hilbert space dimension scales exponentially with system size, exact diagonalization (ED) is practical only in very limited cases. Although there are many approximate methods such as density functional theory \cite{Hohenberg1964,Kohn1965}, density matrix renormalization group \cite{White1992,Verstraete2023}, dynamical mean field theory \cite{Georges1996}, or quantum Monte Carlo \cite{Prokof'ev1996,Suzuki1977,OFS2002,Alet2005}, each method is highly successful for certain problems but has difficulty for others. It is therefore still worthwhile to pursue new quantum many-body methods.

\begin{figure*}
	\centering
	\includegraphics{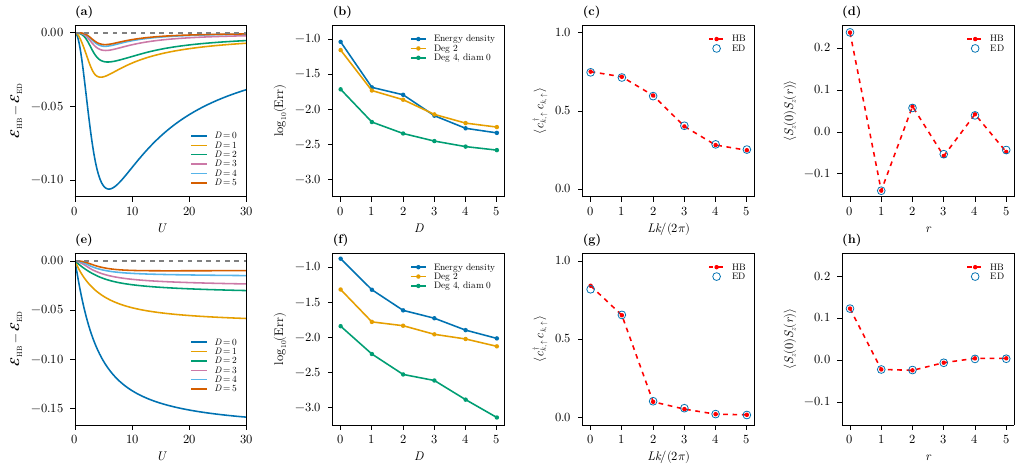}
	\caption{Comparison of HB and ED for the 1D Hubbard model with $L = 10$ sites and hopping parameter $t = 1$. \textbf{(a)}-\textbf{(d)} Show results for half filling while \textbf{(e)}-\textbf{(h)} show results for quarter filling. The interaction strength $U$ is varied in \textbf{(a)} and \textbf{(e)}, but is fixed at $U = 10$ for \textbf{(b)}-\textbf{(d)} and \textbf{(f)}-\textbf{(h)}. $\mathcal{E}_{\text{HB}}$ and $\mathcal{E}_{\text{ED}}$ are the ground state energy densities computed using HB and ED, respectively. \textbf{(a)} and \textbf{(e)} $\mathcal{E}_{\text{HB}} - \mathcal{E}_{\text{ED}}$ as a function of $U$ for varying values of the diameter parameter $D$, which controls the size of $\mathcal{P}$. \textbf{(b)} and \textbf{(f)} The base $10$ logarithm of the error for various ground state quantities as a function of $D$. The quantity indicated by the thick blue lines is the ground state energy density. The other two lines indicate the maximal error across all EVs for operators in $\tilde{\mathcal{Q}}$ [see \cref{eq:tilde-Q}] of support at most $2$ and the indicated degree and diameter. Additional EV error results are shown \cref{fig:correlations}. \textbf{(c)} and \textbf{(g)} The Fermi surface with interactions using $D = 5$. \textbf{(d)} and \textbf{(h)} Spin correlations as a function of distance using $D = 5$. The oscillation in \textbf{(d)} reveals the antiferromagnetic correlations in the ground state at half filling.}
	\label{fig:10-sites}
\end{figure*}

One promising approach that has yet to become commonly used is semidefinite relaxation. The basic idea, which can be traced back at least to Coleman \cite{Coleman1963} is that for a many-body Hamiltonian written in terms of some operator algebra (e.g., fermionic, bosonic, or spin) the ground state energy can often be expressed as a linear combination of a small number of expectation values (EVs). If one minimizes the energy directly over those EVs subject to a relaxed (i.e., weakened) form of the constraint requiring that the EVs are compatible with a physical many-body state, then one will always achieve a lower bound on the true ground state energy. This minimization problem naturally takes the form of a semidefinite program (SDP), which is a type of convex optimization problem for which efficient polynomial scaling numerical algorithms are known \cite{Vandenberghe1996,Boyd2004}. Semidefinite relaxation methods have been studied in the context of quantum chemistry \cite{Mazziotti2020,Mazziotti2023}, condensed matter theory \cite{Barthel2012,Baumgratz2012,Haim2020,Han2020a,Lin2022,Wang2024,Khoo2024,Gao2024}, and high energy theory \cite{Han2020,Berenstein2021,Berenstein2024}.

One consideration that has not yet received sufficient attention is the role of symmetry in semidefinite relaxation. In many earlier works, symmetries such as particle number, spatial translation, or time translation were used to relate EVs or set EVs to zero, thereby reducing the number of optimization variables \cite{Mazziotti2020,Mazziotti2023,Barthel2012,Baumgratz2012,Haim2020,Lin2022,Wang2024,Khoo2024,Gao2024,Han2020,Han2020a,Berenstein2021,Berenstein2024}. In some of these papers, symmetries were additionally used to compress the positive semidefiniteness constraints. Nonetheless, a general and systematic approach to symmetry is still missing.

In this Letter, we describe a semidefinite relaxation method which approximates properties of the ground state of a Hamiltonian subject to Hermitian linear constraints. These constraints can be chosen to project the Hamiltonian into a desired subspace. We call this method Hamiltonian bootstrap (HB) in reference to earlier computational methods based on semidefinite programming which bear the moniker ``bootstrap" \cite{El-Showk2012,Paulos2017-I,Paulos2017-II,Lin2020,Kazakov2003,Chester2023}.

We give a general treatment of symmetry in HB considering unitary, antiunitary, discrete, and continuous symmetries. We use these symmetries to reduce the number of optimization variables and to compress the positive semidefiniteness constraints. This reduces the required computational resources without affecting the accuracy of the results. We derive the method in two pictures related by Lagrangian duality and prove that strong duality holds under mild assumptions.

Finally, we demonstrate HB on the 1D Hubbard model \cite{Gutzwiller1963,Kanamori1963,Hubbard1963,Essler2009}. For a system of $10$ sites, we find quantitative agreement with ED at both half filling and quarter filling. Additionally, for a system of $100$ sites, we find quantitative agreement of the ground state energy density at half filling with the exact infinite system size result from the Bethe ansatz \cite{Lieb1968,Lieb2003}. In these examples, we are able to reduce the sizes of the SDPs by several orders of magnitude using symmetry. Although we only demonstrate HB with a 1D fermionic model, we emphasize that HB can be applied to systems with other degrees of freedom and in higher dimensions.

\emph{Expectation picture}---We now give our first formulation of HB. Let $V$ be a Hilbert space of quantum states and let $\mathcal{L}$ and $\mathcal{H}$ be the spaces of linear operators and Hermitian linear operators on $V$, respectively. We assume that $V$ is finite-dimensional for simplicity, though many of our arguments can be generalized to infinite-dimensional systems. We are given a Hamiltonian $H \in \mathcal{H}$ and a finite set $C \subset \mathcal{H}$ of constraint operators. Our goal is to lower bound the energy
\begin{equation}\label{eq:define-E_0}
\begin{split}
&E_0 = \min_{\rho\succeq 0} \text{tr}(H\rho)\\
&\text{s.t. } \text{tr}(\rho) = 1 \text{ and } \text{tr}(C\rho) = 0 \text{ for all } C \in \mathcal{C}.
\end{split}
\end{equation}
In order to ensure that $E_0$ is well defined, we assume there is some density operator $\rho$ with $\text{tr}(C \rho) = 0$ for all $C \in \mathcal{C}$. If $\mathcal{C}$ is empty, $E_0$ is simply the ground state energy of $H$. Similarly, if $\mathcal{C}$ is nonempty and consists of positive semidefinite (PSD) operators, $E_0$ is the ground state energy of the projection of $H$ into the joint nullspace of the constraint operators.

Consider some $\rho \in \mathcal{H}$. Note that $\rho \succeq 0$ is equivalent to $\text{tr}(X\rho) \geq 0$ for all $X \succeq 0$. Furthermore, any $X \succeq 0$ can be written as a finite sum of Hermitian squares $X = \sum_j \mathcal{O}_j^\dagger \mathcal{O}_j$, where each $\mathcal{O}_j \in \mathcal{L}$. As a result, for any linearly independent set $\mathcal{P} \subset \mathcal{L}$, we can relax the constraint $\rho \succeq 0$ to $\text{tr}(\mathcal{O}^\dagger \mathcal{O}\rho) \geq 0$ for all $\mathcal{O} \in \text{Span}(\mathcal{P})$. The resulting optimization problem is
\begin{equation}\label{eq:tilde-E-P}
\begin{split}
&E_\mathcal{P} = \min_{\rho \in \mathcal{H}} \text{tr}(H\rho)\\
&\text{s.t. } \text{tr}(\mathcal{O}^\dagger \mathcal{O} \rho) \geq 0 \text{ for all } \mathcal{O} \in \text{Span}(\mathcal{P}),\\
&\text{tr}(\rho) = 1, \text{ and } \text{tr}(C\rho) = 0 \text{ for all } C \in \mathcal{C},
\end{split}
\end{equation}
and we have $E_\mathcal{P} \leq E_0$. The set $\mathcal{P}$ serves as the convergence parameter for this approximation, and if $\mathcal{P}$ is a basis for $\mathcal{L}$ then $E_\mathcal{P} = E_0$.

To introduce symmetry, suppose that a compact Lie group $G$ acts on $V$ by a unitary corepresentation $U^V$ \cite{Wigner1959,Dimmock1963,Newmarch1982}, so that each operator $U^V_g$ is either unitary or antiunitary. We assume that $H$ and all constraint operators are invariant under conjugation by $U^V$ and that $\text{Span}(\mathcal{P})$ is closed under conjugation by $U^V$.

Let $G_u$ ($G_a$) be the subgroup (coset) of all $g \in G$ for which $U^V_g$ is unitary (antiunitary). We define a linear representation $U^\mathcal{L}$ of $G$ on $\mathcal{L}$ by
\begin{equation}
U^\mathcal{L}_g(X) = \begin{cases}
U^V_g X (U^V_g)^\dagger & \text{for } g \in G_u\\
U^V_g X^\dagger (U^V_g)^\dagger & \text{for } g \in G_a
\end{cases}
\end{equation}
for all $X \in \mathcal{L}$. We note that $H$ and all constraint operators are invariant under $U^\mathcal{L}$. If $G_a$ is nonempty, we assume that $\text{Span}(\mathcal{P})$ is closed under adjoints so that $\text{Span}(\mathcal{P})$ is also closed under $U^\mathcal{L}$. By working with the linear representation $U^\mathcal{L}$ rather than the corepresentation $U^V$ we can circumvent some of the complications of antilinear symmetries.

Since Haar's theorem guarantees the existence of a left-invariant probability measure $\mu$ on $G$ \cite{Halmos1976}, we can define the averaging map
\begin{equation}\label{eq:averaging-map}
A(X) = \int_{g \in G} U^\mathcal{L}_g(X) d\mu(g)
\end{equation}
for all $X \in \mathcal{L}$. Note that for any admissible solution $\rho$ of \cref{eq:tilde-E-P} with energy $E$, $U^\mathcal{L}_g(\rho)$ is also an admissible solution with energy $E$ for all $g \in G$, and therefore $A(\rho)$ is as well. Since $A(\rho)$ is invariant under $U^\mathcal{L}$, we can add the constraint
\begin{equation}\label{eq:rho-U}
U^\mathcal{L}_g(\rho) = \rho \text{ for all } g \in G
\end{equation}
to \cref{eq:tilde-E-P} without changing the value of $E_\mathcal{P}$.

We now seek to simplify our optimization problem. We take $\mathcal{P} = \{p_1, \dots, p_m\}$ and introduce a Hermitian basis $\mathcal{Q} = \{q_1, \dots, q_n\}$ for $\text{Span}(\{A(p_j^\dagger p_k) | 1 \leq j, k \leq m\})$. This allows us to define the Hermitian structure constant matrices $\Gamma^l$ for $1 \leq l \leq n$ by
\begin{equation}\label{eq:structure-constants}
A(p_j^\dagger p_k) = \sum_{l=1}^n \Gamma^l_{j, k} q_l \text{ for all } 1 \leq j, k \leq m.
\end{equation}

Now suppose $\rho$ satisfies \cref{eq:rho-U} and let $\mathcal{O} = \sum_{j=1}^m z_j p_j$ for some vector $z \in \mathbb{C}^m$. We note that $\text{tr}(\mathcal{O}^\dagger \mathcal{O}\rho) = \text{tr}(A(\mathcal{O}^\dagger \mathcal{O}) \rho)$. Additionally, \cref{eq:structure-constants} implies $A(\mathcal{O}^\dagger \mathcal{O}) = \sum_{l=1}^n (z^\dagger \Gamma^l z) q_l$. Together, these imply
\begin{equation}\label{eq:trace-z-tilde-f}
\text{tr}(\mathcal{O}^\dagger \mathcal{O} \rho) = z^\dagger f(\rho)z
\end{equation}
where $f(\rho) = \sum_{l=1}^n \text{tr}(q_l \rho) \Gamma^l$. It follows that $\text{tr}(\mathcal{O}^\dagger \mathcal{O} \rho) \geq 0$ for all $\mathcal{O} \in \text{Span}(\mathcal{P})$ is equivalent to $f(\rho) \succeq 0$.

We assume that $H$, the identity operator $I$, and all constraint operators are contained in $\text{Span}(\{p_j^\dagger p_k | 1 \leq j, k \leq m\})$. Since these operators are invariant under $U^\mathcal{L}$, they are also contained in $\text{Span}(\mathcal{Q})$ so we can expand them as
\begin{equation}\label{eq:H-I-C-expansion}
H = \sum_{l=1}^n H_l q_l,\quad I = \sum_{l=1}^n I_l q_l,\quad C = \sum_{l=1}^n C_l q_l
\end{equation}
for all $C \in \mathcal{C}$. It follows that $E_\mathcal{P}$ is given by the SDP \footnote{We note that \cref{eq:dual-original} does not exactly match the standard definition of an SDP \cite{Vandenberghe1996,Boyd2004}. Nonetheless, we use the term SDP to refer to any convex optimization problem involving extremization of a linear function of real scalar and Hermitian matrix variables subject to affine and positive semidefiniteness constraints. In this work, we solve SDPs using the Mosek Optimizer API for Julia \cite{Mosek}.}
\begin{equation}\label{eq:dual-original}
\begin{split}
&E_\mathcal{P} = \min_x \sum_{l=1}^n H_l x_l \text{ s.t. } \sum_{l=1}^n x_l \Gamma^l \succeq 0,\\
&\sum_{l=1}^n I_l x_l = 1, \text{ and } \sum_{l=1}^n C_l x_l = 0 \text{ for all } C \in \mathcal{C}
\end{split}
\end{equation}
where $x \in \mathbb{R}^n$ is the vector with $x_l = \text{tr}(q_l \rho)$.

So far, we have used symmetry to reduce the number $n$ of optimization variables by including the averaging map $A$ in the definition of $f(\rho)$. We now show that symmetry can additionally be used to compress the positive semidefiniteness constraint $f(\rho) \succeq 0$. We define a linear representation $U^\mathcal{P}$ of $G_u$ on $\mathbb{C}^m$ by
\begin{equation}\label{eq:define-U-P}
U^\mathcal{L}_g(p_j) = \sum_{k=1}^m (U^\mathcal{P}_g)_{k,j} p_k \text{ for all } g \in G_u
\end{equation}
and $1 \leq j \leq m$. We note that for any $g \in G_u$, the mapping $\mathcal{O} \mapsto U^\mathcal{L}_g(\mathcal{O})$ is equivalent to $z \mapsto U^\mathcal{P}_g z$. Since the left-hand side of \cref{eq:trace-z-tilde-f} is invariant under this mapping, $f(\rho)$ must be invariant under conjugation by $U^\mathcal{P}$ \footnote{This argument is restricted to $g \in G_u$ since if $g \in G_a$ then under $\mathcal{O} \mapsto U^\mathcal{L}_g(\mathcal{O})$ we have $\text{tr}(\mathcal{O}^\dagger \mathcal{O}\rho) \mapsto \text{tr}(\mathcal{O} \mathcal{O}^\dagger \rho)$. That is to say, the left-hand side of \cref{eq:trace-z-tilde-f} is only invariant under the mapping for $g \in G_u$.}.

We assume that $U^\mathcal{P}$ is a unitary representation so that it can be decomposed into irreducible representations (irreps) of $G_u$ as $\mathcal{V}^\dagger U^{\mathcal{P}}_g \mathcal{V} = \oplus_{\lambda=1}^\Lambda (I_{m_\lambda} \otimes U^\lambda_g)$ for all $g \in G_u$ and some unitary matrix $\mathcal{V}$ \footnote{Since $G_u \subset G$ is compact, $U^\mathcal{P}$ is completely reducible even if we do not assume it is unitary. However, we need to assume that $U^\mathcal{P}$ is unitary in order to ensure that the basis transformation $\mathcal{V}$ is unitary, which is needed to ensure that the $\Gamma^{l,\lambda}$ matrices in \cref{eq:Gamma-l-lambda} are Hermitian.}. Here, $U^1, \dots, U^\Lambda$ are nonisomorphic unitary irreps of $G_u$, $m_1, \dots, m_\Lambda$ are the corresponding multiplicities, and $I_\alpha$ denotes the identity matrix of dimension $\alpha$. By Schur's lemma, a square matrix $Z$ of dimension $m$ is invariant under conjugation by $U^\mathcal{P}$ if and only if
\begin{equation}\label{eq:Hermitian-decomposition}
\mathcal{V}^\dagger Z \mathcal{V} = \oplus_{\lambda=1}^\Lambda (Z_\lambda \otimes I_{m'_\lambda})
\end{equation}
for some matrices $Z_1, \dots, Z_\Lambda$, where the dimension of $Z_\lambda$ is $m_\lambda$ and the dimension of $U^\lambda$ is $m'_\lambda$.

Since $f(\rho)$ is invariant under conjugation by $U^\mathcal{P}$ it can be decomposed as in \cref{eq:Hermitian-decomposition}. Let the columns of $\mathcal{V}$ corresponding to the $j$th copy of irrep $U^\lambda$ in $U^\mathcal{P}$ be denoted $\mathcal{V}^{j,\lambda}$ for all $1 \leq j \leq m_\lambda$ and $1 \leq \lambda \leq \Lambda$, so that $\mathcal{V}^{j,\lambda}$ is an $m \times m'_\lambda$ matrix. We define the Hermitian symmetrized structure constant matrices $\Gamma^{l,\lambda}$ for $1 \leq l \leq n$ by
\begin{equation}\label{eq:Gamma-l-lambda}
\Gamma^{l,\lambda}_{j, k} = \text{tr}\left((\mathcal{V}^{j,\lambda})^\dagger \Gamma^l \mathcal{V}^{k,\lambda}\right) \text{ for all } 1 \leq j, k \leq m_\lambda.
\end{equation}
We then have $\mathcal{V}^\dagger f(\rho) \mathcal{V} = \oplus_{\lambda=1}^\Lambda (f_\lambda(\rho) \otimes I_{m'_\lambda}/m'_\lambda)$ where $f_\lambda(\rho) = \sum_{l=1}^n \text{tr}(q_l \rho) \Gamma^{l,\lambda}$ for all $1 \leq \lambda \leq \Lambda$. It follows that $f(\rho) \succeq 0$ if and only if $f_1(\rho), \dots, f_\Lambda(\rho) \succeq 0$, and therefore $E_\mathcal{P}$ is given by the SDP
\begin{equation}\label{eq:dual-symmetrized}
\begin{split}
&E_\mathcal{P} = \min_x \sum_{l=1}^n H_l x_l\\
&\text{s.t. } \sum_{l=1}^n x_l \Gamma^{l,\lambda} \succeq 0 \text{ for all } 1 \leq \lambda \leq \Lambda,\\
&\sum_{l=1}^n I_l x_l = 1, \text{ and } \sum_{l=1}^n C_l x_l = 0 \text{ for all } C \in \mathcal{C}.
\end{split}
\end{equation}
Since the optimization variables $x_l$ are ground state EVs, we refer to \cref{eq:dual-symmetrized} and its derivation as the expectation picture. Furthermore, we note that any EV $\text{tr}(q\rho)$ for $q$ in
\begin{equation}\label{eq:tilde-Q}
\tilde{\mathcal{Q}} = \{p_j^\dagger p_k | 1 \leq j, k \leq m\}
\end{equation}
can be expressed as a linear combination of the $x_l$ variables using symmetry.

\begin{figure}
	\centering
	\includegraphics{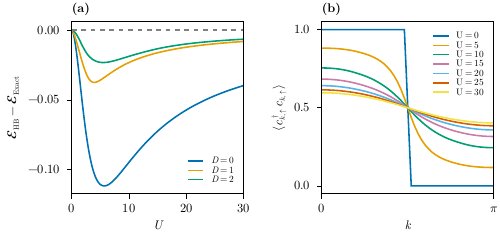}
	\caption{HB for the 1D Hubbard model with $L = 100$ sites and hopping parameter $t = 1$ at half filling. $\mathcal{E}_{\text{HB}}$ is the ground state energy density computed using HB while $\mathcal{E}_{\text{Exact}}$ is the exact ground state energy density in the thermodynamic limit, as computed in Refs. \cite{Lieb1968,Lieb2003} using the Bethe ansatz. \textbf{(a)} $\mathcal{E}_{\text{HB}} - \mathcal{E}_{\text{Exact}}$ as a function of the interaction strength $U$ for several values of the diameter parameter $D$, which controls the size of $\mathcal{P}$. \textbf{(b)} The Fermi surface using $D = 2$ and several values of $U$.}
	\label{fig:100-sites}
\end{figure}

\emph{Sum of squares picture}---We now give our second formulation of HB. Suppose that
\begin{equation}\label{eq:primal-start}
H - EI - \sum_{C \in \mathcal{C}} \gamma_C C \succeq 0
\end{equation}
holds for some real value $E$ and real vector $\gamma$. By tracing against an optimal solution of \cref{eq:define-E_0}, we see that $E \leq E_0$. We therefore define $E'_0$ to be the largest value $E$ such that \cref{eq:primal-start} holds for some $\gamma$. We have $E'_0 \leq E_0$, and we will see later that $E'_0 = E_0$ under mild assumptions.

Since every PSD operator can be written as a sum of Hermitian squares, and additionally $H$, $I$, and all constraint operators are invariant under the averaging map $A$,
\cref{eq:primal-start} is equivalent to
\begin{equation}\label{eq:primal-sum-of-squares}
H = EI + \sum_{C \in \mathcal{C}} \gamma_C C + \sum_j A(\mathcal{O}^\dagger_j \mathcal{O}_j)
\end{equation}
for some finite collection of operators $\mathcal{O}_j \in \mathcal{L}$. We now define $E'_\mathcal{P}$ to be the largest value $E$ such that \cref{eq:primal-sum-of-squares} holds for some $\gamma$ and operators $\mathcal{O}_j \in \text{Span}(\mathcal{P})$, and note that $E'_\mathcal{P} \leq E'_0$. The set $\mathcal{P}$ serves as the convergence parameter for this approximation, and if $\mathcal{P}$ is a basis for $\mathcal{L}$ then $E'_\mathcal{P}= E'_0$. By tracing \cref{eq:primal-sum-of-squares} against an optimal solution $\rho$ of \cref{eq:define-E_0,eq:rho-U}, we see that
\begin{equation}\label{eq:equality-condition}
E'_\mathcal{P} = E_0 \text{ if and only if } \text{tr}(\mathcal{O}^\dagger_j \mathcal{O}_j \rho) = 0 \text{ for all } j.
\end{equation}

Suppose $\mathcal{O}_j = \sum_{k=1}^m z^{(j)}_k p_k$ for some vectors $z^{(j)} \in \mathbb{C}^m$. By \cref{eq:structure-constants}, we have
\begin{equation}\label{eq:A-O-F-Z}
\sum_j A(\mathcal{O}^\dagger_j \mathcal{O}_j) = F(Z)
\end{equation}
where $Z = \sum_j z^{(j)} (z^{(j)})^\dagger$ and $F(Z) = \sum_{l=1}^n \text{tr}(\Gamma^l Z)q_l$. Since $Z$ can be any PSD matrix of dimension $m$, $E'_\mathcal{P}$ is given by the SDP
\begin{equation}\label{eq:primal-original}
\begin{split}
&E'_{\mathcal{P}} = \max_{E, \gamma} \max_{Z\succeq 0} E\\
&\text{s.t. } H_l = E I_l + \sum_{C \in \mathcal{C}} \gamma_C C_l + \text{tr}(\Gamma^l Z)\\
&\text{for all } 1 \leq l \leq n,
\end{split}
\end{equation}
where we have used the expansions in \cref{eq:H-I-C-expansion}.

So far, we have used symmetry to reduce the number $n$ of constraint equations in \cref{eq:primal-original} by including the averaging map $A$ in the definition of $F(Z)$. We now show that symmetry can additionally be used to compress the positive semidefiniteness constraint $Z \succeq 0$. We define the averaging map $A^\mathcal{P}$ for the representation $U^\mathcal{P}$ in \cref{eq:define-U-P} by
\begin{equation}
A^\mathcal{P}(M) = \int_{g \in G_u} U^\mathcal{P}_g M (U^\mathcal{P}_g)^\dagger d\mu(g)
\end{equation}
for any square matrix $M$ of dimension $m$. For any $g \in G_u$, the mapping $z^{(j)} \mapsto U^\mathcal{P}_g z^{(j)}$ sends $\mathcal{O}_j \mapsto U^\mathcal{L}_g(\mathcal{O}_j)$, and therefore leaves $F(Z)$ invariant by \cref{eq:A-O-F-Z}. On the other hand, this mapping takes $Z \mapsto U^\mathcal{P}_g Z (U^\mathcal{P}_g)^\dagger$. It follows that if $(E, \gamma, Z)$ is an admissible solution of \cref{eq:primal-original}, then $(E, \gamma, U^\mathcal{P}_g Z (U^\mathcal{P}_g)^\dagger)$ is also an admissible solution for all $g \in G_u$, and therefore $(E, \gamma, A^\mathcal{P}(Z))$ is as well. Since $A^\mathcal{P}(Z)$ is invariant under conjugation by $U^\mathcal{P}$, we can add the constraint
\begin{equation}\label{eq:Z-U}
U^\mathcal{P}_g Z (U^\mathcal{P}_g)^\dagger = Z \text{ for all } g \in G_u
\end{equation}
to \cref{eq:primal-original} without changing the value of $E'_\mathcal{P}$. By expanding $Z$ as in \cref{eq:Hermitian-decomposition}, we see that $Z \succeq 0$ if and only if $Z_1, \dots, Z_\Lambda \succeq 0$. Furthermore, $\text{tr}(\Gamma^l Z) = \sum_{\lambda=1}^\Lambda \text{tr}(\Gamma^{l,\lambda} Z_\lambda)$ where $\Gamma^{l,\lambda}$ is given by \cref{eq:Gamma-l-lambda}. Finally, $E'_\mathcal{P}$ is given by the SDP
\begin{equation}\label{eq:primal-symmetrized}
\begin{split}
&E'_{\mathcal{P}} = \max_{E, \gamma} \max_{Z_1,\dots,Z_\Lambda\succeq 0} E\\
&\text{s.t. } H_l = E I_l + \sum_{C \in \mathcal{C}} \gamma_C C_l + \sum_{\lambda=1}^\Lambda \text{tr}(\Gamma^{l,\lambda} Z_\lambda)\\
&\text{for all } 1 \leq l \leq n.
\end{split}
\end{equation}
Since any optimal solution of this SDP corresponds to an identity of the form in \cref{eq:primal-sum-of-squares}, we refer to \cref{eq:primal-symmetrized} and its derivation as the sum of squares picture.

\emph{Duality}---Although we gave largely independent derivations of the SDPs in \cref{eq:dual-symmetrized,eq:primal-symmetrized}, they are related by Lagrangian duality \cite{Vandenberghe1996,Boyd2004}. It follows from the weak duality theorem that $E'_\mathcal{P} \leq E_\mathcal{P}$ always holds. It is not immediately guaranteed that strong duality (i.e., $E'_\mathcal{P} = E_{\mathcal{P}}$) holds. However, we prove in \cref{app:strong-duality} that strong duality does hold whenever the intersection of the nullspaces of the constraint operators contains a nonzero vector. By taking $\mathcal{P}$ to be a basis for $\mathcal{L}$, it follows that $E'_0 = E_0$ whenever this condition is satisfied.

\emph{Example}---We now apply HB to the 1D Hubbard model \cite{Gutzwiller1963,Kanamori1963,Hubbard1963,Essler2009} with an even number $L$ of sites, periodic boundary conditions, hopping parameter $t \geq 0$, interaction parameter $U \geq 0$, and chemical potential $\mu = \frac{U}{2}$ (see \cref{app:hubbard} for details). This system has discrete unitary symmetries (translation and inversion), a discrete antiunitary symmetry (complex conjugation), and continuous unitary symmetries (spin and eta pairing \cite{Yang1989,Yang1990}). We consider the overall ground state which has half filling by taking $\mathcal{C} = \{\}$ and the ground state at quarter filling by taking $\mathcal{C} = \{C_0, C_0^2\}$ where $C_0 = N - LI/2$ and where $N$ is the number operator \footnote{Although the constraint $C_0$ is logically redundant with $C_0^2$, we find better numerical performance when both constraints are supplied.}. When studying the overall ground state, we use all the symmetries described above. However, when studying the quarter filling ground state, we must elide the $x$ and $y$ components of eta pairing from the list of Lie algebra generators.

\cref{fig:10-sites,fig:correlations} compare HB with ED for $L = 10$ and $t = 1$ at half filling and quarter filling. For the HB results, we take $\mathcal{P}$ to be a finite set of ordered Majorana operator products. Specifically, for some integer $D \geq 0$, $\mathcal{P}$ contains all degree $0$ and $1$ products, all degree $2$ products of diameter at most $D$, and all degree $3$ products of diameter at most $D$ and support at most $2$ (see \cref{app:hubbard} for details). The largest set $\mathcal{P}$ used for $L = 10$ has $D = 5$ and contains $m = 3021$ elements.

At half filling, we see that the error in the energy density decreases monotonically in $D$ for all $U$, and vanishes at $U = 0$ and $U \to \infty$. This vanishing error can be understood in the sum of squares picture, since the condition in \cref{eq:equality-condition} can be satisfied for any $D$ when either $U = 0$, $t = 1$ or $U = 1$, $t = 0$. The energy density and all EVs of degree at most $4$ and support at most $2$ are accurate for $D = L/2$. Additionally the energy density, all degree $2$ EVs, and all degree $4$ EVs of diameter $0$ are accurate even for small $D$.

The results at quarter filling are similar, but it is interesting to note that the ground state at quarter filling has a symmetry enforced fourfold degeneracy while the ground state at half filling is nondegenerate. We therefore use the density operator $\rho_0 = P_0/\text{tr}(P_0)$ in ED, where $P_0$ is the projector onto the ground state subspace.

\begin{figure}
	\centering
	\includegraphics{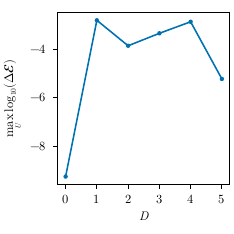}
	\caption{Comparison of the HB energy density for the 1D Hubbard model with $L = 10$ sites and hopping parameter $t = 1$ at half filling for two different choices of $\mathcal{P}$ as a function of the diameter parameter $D$. Both choices of $\mathcal{P}$ consist of ordered Majorana operator products. One choice contains all degree $1$ products and degree $3$ products of diameter at most $D$ and support at most $2$. The other choice contains the aforementioned operators along with all degree $0$ products and degree $2$ products of diameter at most $D$. For each $D$, we show the maximum base $10$ logarithm of the energy density difference across all values of $U$ between $0$ and $30$ which are an integer multiple of $\frac{1}{10}$.}
	\label{fig:basis-sets}
\end{figure}

\cref{fig:100-sites} shows HB results for $L = 100$ and $t = 1$ at half filling. We again take $\mathcal{P}$ to consist of ordered Majorana products. However, for efficiency we take $\mathcal{P}$ to consist only of degree $1$ products and degree $3$ products of diameter at most $D$ and support at most $2$. This is motivated by \cref{fig:basis-sets} which demonstrates that removing the degree $0$ and $2$ operators leaves the energy density almost exactly the same for $L = 10$. The largest set $\mathcal{P}$ used for $L = 100$ has $D = 2$ and contains $m = 10400$ elements.

\cref{fig:100-sites}\textbf{(a)} compares HB with the exact infinite size energy density from the Bethe ansatz. Since the error in the energy density as a function of $U$ and $D$ appears similar to that in \cref{fig:10-sites}\textbf{(a)}, we expect degree $2$ EVs and degree $4$ EVs of diameter $0$ to be accurate even for small $D$. Indeed, the Fermi surface under interaction with $D = 2$ shown in \cref{fig:100-sites}\textbf{(b)} flattens as $U$ increases, as one would expect.

\cref{app:resources} discusses the scaling of the Hubbard model SDPs as a function of $L$ both with and without symmetrization. As illustrated in \cref{fig:scaling}, the use of symmetry can reduce the effective number of variables and constraints by many orders of magnitude.

\emph{Discussion}---In the worst case, HB requires $m \sim \text{dim}(V)$ constraint operators in order to obtain an accurate ground state energy. In this case, the method scales exponentially with system size just like ED. However, physical Hamiltonians are not generic. In particular, they are often low degree and sparse (i.e., they can be written in terms of a small number of simple operators) and highly symmetric. Since HB is able to make use of these properties, we expect that HB will be practically useful with polynomial resources for many problems in quantum chemistry, condensed matter theory, and high energy theory.

\emph{Acknowledgements}---We thank Jonah Herzog-Arbeitman, Ross Dempsey, Benjamin S{\o}gaard, Yves H. Kwan, Biao Lian, Nicolas Regnault, B. Andrei Bernevig, Nisarg Chadha, and Eslam Khalaf for numerous helpful discussions. This work was supported by the Princeton Center for Complex Materials, a National Science Foundation Materials Research Science and Engineering Center under Grant No. NSF DMR-2011750. Additionally, this work was supported by the National Science Foundation under Grant No. NSF DMR-2141966.

\bibliography{bibliography}
\appendix

\section{Strong duality}\label{app:strong-duality}
We now present two criteria under which the SDPs in \cref{eq:dual-symmetrized,eq:primal-symmetrized} satisfy strong duality (i.e., $E'_\mathcal{P} = E_\mathcal{P}$). The first criterion states that for all $\epsilon > 0$, there is a strictly positive definite operator $\rho_\epsilon \in \mathcal{L}$ such that $\text{tr}(\rho_\epsilon) = 1$ and $|\text{tr}(C\rho_\epsilon)| \leq \epsilon$ for all $C \in \mathcal{C}$. With this assumption, the optimization problem
\begin{equation}
\begin{split}
&\min_x \sum_{l=1}^n H_l x_l\\
&\text{s.t. } \sum_{l=1}^n x_l \Gamma^{l,\lambda} \succeq 0 \text{ for all } 1 \leq \lambda \leq \Lambda,\\
&\sum_{l=1}^n I_l x_l = 1, \text{ and } \left|\sum_{l=1}^n C_l x_l\right| \leq \epsilon \text{ for all } C \in \mathcal{C}
\end{split}
\end{equation}
satisfies Slater's condition \cite{Boyd2004} and therefore has strong duality for all $\epsilon > 0$. Since the duality gap (i.e., the difference between the primal and dual optimal objective values) is continuous in $\epsilon$, we can take $\epsilon \to 0$ and conclude that \cref{eq:dual-symmetrized,eq:primal-symmetrized} have strong duality.

The second criterion states that the intersection of the nullspaces of the constraint operators contains a nonzero vector. With this assumption, the operator $\tilde{\mathcal{C}} = \sum_{C \in \mathcal{C}} C^2$ has $0$ as its lowest eigenvalue. We then define $\rho_\delta = e^{-\delta \tilde{\mathcal{C}}}/\text{tr}(e^{-\delta \tilde{\mathcal{C}}})$ for all $\delta \geq 0$. It is clear that $\rho_\delta$ is strictly positive definite and has $\text{tr}(\rho_\delta) = 1$. Furthermore, since $\lim_{\delta\to \infty} \text{tr}(\tilde{\mathcal{C}}\rho_\delta) = 0$, it follows that for any $\epsilon > 0$ there is a $\delta$ large enough such that $\text{tr}(\tilde{\mathcal{C}} \rho_\delta) \leq \epsilon^2$. For this choice of $\delta$, we then have $\text{tr}(C\rho_\delta)^2 \leq \text{tr}(C^2 \rho_\delta) \leq \text{tr}(\tilde{\mathcal{C}} \rho_\delta) \leq \epsilon^2$ and so $|\text{tr}(C\rho_\delta)| \leq \epsilon$, for all $C \in \mathcal{C}$. As a result, the first criterion is satisfied so the SDPs have strong duality.

\begin{figure}
	\centering
	\includegraphics{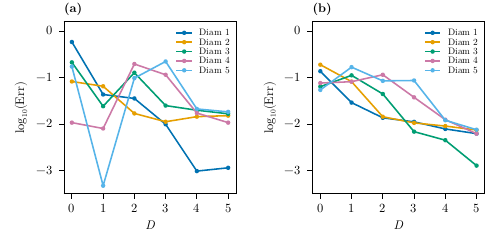}
	\caption{Additional EV error results for HB relative to ED for the 1D Hubbard model with $L = 10$ sites, hopping parameter $t = 1$, and interaction strength $U = 10$ at \textbf{(a)} half filling and \textbf{(b)} quarter filling. The results are shown as a function of the diameter parameter $D$, which controls the size of $\mathcal{P}$. Each line indicates the base $10$ logarithm of the maximal error across all EVs for operators in $\tilde{\mathcal{Q}}$ [see \cref{eq:tilde-Q}] of support at most $2$, degree $4$, and the indicated diameter. Similar results for other quantities are shown in \cref{fig:10-sites}\textbf{(b)} and \textbf{(f)}.}
	\label{fig:correlations}
\end{figure}

\section{1D Hubbard model}\label{app:hubbard}
We now give details of the application of HB to the 1D Hubbard model \cite{Gutzwiller1963,Kanamori1963,Hubbard1963,Essler2009}. We consider a system with $L$ sites and periodic boundary conditions and denote the fermionic annihilation operator at site $r \in \{0, \dots, L-1\}$ with spin $s \in \{\uparrow, \downarrow\}$ by $c_{r,s}$. Because of the periodic boundary conditions, we will allow the site index $r$ to range over all integers with the understanding that $c_{r,s} = c_{r+L,s}$ for all $r$. These operators satisfy the anticommutation relations $\{c_{r,s}, c^\dagger_{r',s'}\} = \delta_{r,r'}\delta_{s,s'}I$ and $\{c_{r,s}, c_{r',s'}\} = 0$ so that the Hilbert space $V$ is a fermionic Fock space with dimension $2^{2L}$. The Hamiltonian takes the form
\begin{equation}
\begin{split}
H &= -t\sum_{r=0}^{L-1} \sum_{s \in \{\uparrow,\downarrow\}} c^\dagger_{r,s} c_{r+1,s} + c^\dagger_{r+1,s} c_{r,s}\\
&+U \sum_{r=0}^{L-1} c^\dagger_{r,\uparrow} c_{r,\uparrow} c^\dagger_{r,\downarrow} c_{r,\downarrow} -\mu \sum_{r=0}^{L-1} \sum_{s \in \{\uparrow,\downarrow\}} c^\dagger_{r,s} c_{r,s}
\end{split}
\end{equation}
for hopping parameter $t$, interaction parameter $U$, and chemical potential $\mu$. We assume for simplicity that $t, U \geq 0$.

The Hamiltonian commutes with discrete symmetry generators
\begin{equation}\label{eq:discrete-symmetries}
\begin{split}
\text{Translation:}&\quad T c^\dagger_{r,s} T^\dagger = c^\dagger_{r+1,s}\\
\text{Inversion:}&\quad \mathcal{I} c^\dagger_{r,s} \mathcal{I}^\dagger = c^\dagger_{-r,s}\\
\text{Complex conjugation:}&\quad \mathcal{K} (\lambda c^\dagger_{r,s}) \mathcal{K}^\dagger = \lambda^* c^\dagger_{r,s}.
\end{split}
\end{equation}
Translation and inversion are unitary symmetries while complex conjugation is antiunitary. Additionally, the Hamiltonian commutes with the spin $\text{su}(2)$ algebra generated by $S_x = \frac{1}{2}(S_+^\dagger + S_+)$, $S_y = \frac{i}{2}(S_+^\dagger - S_+)$, and $S_z = -i[S_x, S_y]$, where
\begin{equation}\label{eq:S_+}
S_+ = \sum_{r=0}^{L-1} c^\dagger_{r,\uparrow} c_{r,\downarrow}.
\end{equation}
Lastly, when $L$ is even and $\mu = \frac{U}{2}$, the Hamiltonian commutes with the eta pairing $\text{su}(2)$ algebra generated by $\eta_x = \frac{1}{2}(\eta_+^\dagger + \eta_+)$, $\eta_y = \frac{i}{2}(\eta_+^\dagger - \eta_+)$, and $\eta_z = -i[\eta_x, \eta_y]$, where \cite{Yang1989,Yang1990}
\begin{equation}\label{eq:eta_+}
\eta_+ = \sum_{r=0}^{L-1} (-1)^r c^\dagger_{r,\uparrow} c^\dagger_{r,\downarrow}.
\end{equation}
It is worth noting that $2\eta_z + LI = N$ where $N = \sum_{r=0}^{L-1} \sum_{s\in\{\uparrow,\downarrow\}} c^\dagger_{r,s} c_{r,s}$ is the number operator.

We assume from here on that $L$ is even and $\mu = \frac{U}{2}$. In the case that $C = \{\}$, we take $G$ to be the group $G_{\text{full}}$ with discrete generators $T$, $\mathcal{I}$, and $\mathcal{K}$ and Lie algebra generators $S_x$, $S_y$, $S_z$, $\eta_x$, $\eta_y$, and $\eta_z$. In the case that $\mathcal{C} = \{C_0, C_0^2\}$ where $C_0 = N - LI/2$, we take $G$ to be the group $G_{\text{partial}}$ with discrete generators $T$, $\mathcal{I}$, $\mathcal{K}$ and Lie algebra generators $S_x$, $S_y$, $S_z$, and $\eta_z$. In either case, we take $U^V$ to be the symmetry representation defined by \cref{eq:discrete-symmetries,eq:S_+,eq:eta_+}.

We define the Majorana operators $\gamma_{r,s,\sigma}$ for sites $r$, spins $s$, and signs $\sigma \in \{+, -\}$ by $\gamma_{r,s,+} = c^\dagger_{r,s} + c_{r,s}$ and $\gamma_{r,s,-} = i(c^\dagger_{r,s} - c_{r,s})$. These operators are Hermitian and satisfy the anticommutation relation $\{\gamma_{r,s,\sigma}, \gamma_{r',s',\sigma'}\} = 2\delta_{r,r'}\delta_{s,s'}\delta_{\sigma,\sigma'}I$. In particular, each Majorana operator squares to the identity.

We choose some ordering on the set of Majorana operators and then define $\mathcal{M}$ to be the set of all products of the form $\gamma_{r_1,s_1,\sigma_1}\dots\gamma_{r_l,s_l,\sigma_l}$ for $l \geq 0$, where the Majorana operators are distinct and ordered. Let $p \in \mathcal{M}$ be a product of $d$ distinct Majorana operators on sites $r_1, \dots, r_d \in \{0, \dots, L-1\}$. We say that the degree of $p$ is $d$, the diameter of $p$ is $\max(\{\text{dist}(r_j, r_k) | 1 \leq j, k \leq d\})$ where $\text{dist}(r, r') = \min(\{\text{mod}(r - r', L), \text{mod}(r' - r, L)\})$, and the support of $p$ is the number of distinct sites among $r_1, \dots, r_d$.

We choose the set $\mathcal{P}$ to be a subset of $\mathcal{M}$ as described in the main text. It follows that each element $q$ of the set $\tilde{\mathcal{Q}}$ defined in \cref{eq:tilde-Q} takes the form $q = s p$ for $s \in \{1, -1\}$ and $p \in \mathcal{M}$. We define the degree, diameter, and support of $q$ to be the same as for $p$.

Since $\text{tr}(p_1^\dagger p_2) = 2^{2L}\delta_{p_1,p_2}$ for all $p_1, p_2 \in \mathcal{M}$, it follows that the representation $U^\mathcal{P}$ in \cref{eq:define-U-P} is unitary. It is worth noting that this Majorana operator construction for the set $\mathcal{P}$ can be applied generally to fermionic systems, not just to the 1D Hubbard model.

Finally, we define the notations used in \cref{fig:10-sites}\textbf{(c)}, \textbf{(d)}, \textbf{(g)}, \textbf{(h)}. The momentum space creation operators are defined by $c^\dagger_{k,s} = \frac{1}{\sqrt{L}} \sum_{r=0}^{L-1} e^{ikr} c^\dagger_{r,s}$ for $k = \frac{2\pi n}{L}$ and integers $n$. The local spin operators are defined by $S_z(r) = \frac{1}{2}(c^\dagger_{r,\uparrow} c_{r,\uparrow} - c^\dagger_{r,\downarrow} c_{r,\downarrow})$ for sites $r$.

\begin{figure}
	\centering
	\includegraphics{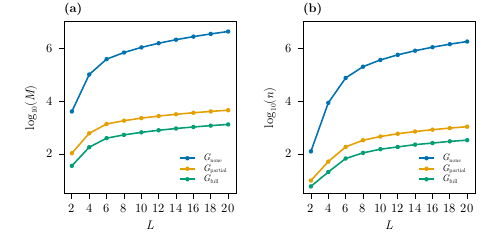}
	\caption{\textbf{(a)} and \textbf{(b)} The symmetrized SDP size parameters $\log_{10}(M)$ and $\log_{10}(n)$, respectively, as a function of $L$ for the 1D Hubbard model using the set $\mathcal{P}$ described in \cref{app:resources}. We evaluate these parameters for the symmetry groups $G_{\text{none}}$, $G_{\text{partial}}$, and $G_{\text{full}}$.}
	\label{fig:scaling}
\end{figure}

\section{Computational savings from symmetrization}\label{app:resources}
In order to demonstrate the practical benefits afforded by symmetry, we illustrate in \cref{fig:scaling} the sizes of the SDPs in \cref{eq:dual-symmetrized,eq:primal-symmetrized} for the 1D Hubbard model with various system sizes and symmetry groups. For simplicity, we choose $\mathcal{P}\subset \mathcal{M}$ to consist of all degree $1$ products and all degree $3$ products of diameter at most $2$ and support at most $2$. We consider the symmetry groups $G_{\text{full}}$ and $G_{\text{partial}}$ defined in \cref{app:hubbard} as well as the trivial group $G_{\text{none}} = \{I\}$. We plot $\log_{10}(n)$ where $n$ is the number of variables (constraint equations) in \cref{eq:dual-symmetrized} [\cref{eq:primal-symmetrized}] and $\log_{10}(M)$ where $M = \sum_{\lambda=1}^\Lambda m_\lambda^2$ is the effective size of the positive semidefiniteness constraints (positive semidefinite variables) in \cref{eq:dual-symmetrized} [\cref{eq:primal-symmetrized}].

We note that symmetries such as translation which generate a group that grows with system size can improve the asymptotic scaling of $n$ and $M$ with $L$. This explains why the curves in \cref{fig:scaling} for $G_{\text{full}}$ and $G_{\text{partial}}$ have substantially lower slopes for large $L$ than that of $G_{\text{none}}$. On the other hand, symmetries such as $\eta_x$, $\eta_y$, and $\eta_z$ which generate a group of fixed size do not improve the asymptotic scaling of $n$ and $M$ with $L$ but nonetheless can significantly reduce $n$ and $M$. This explains why the curves in \cref{fig:scaling} for $G_{\text{full}}$ and $G_{\text{partial}}$ have nearly identical slopes for large $L$ along with a significant offset.

\end{document}